# Physics adapted generative AI for metal insulator transition materials under label scarcity

Gourab Datta[1,2], Sarah Sharif[1,2], Zhisheng Shi[2], Yaser Banad[1,2]*

[1]Intelligent Neuromorphic and Quantum Understanding for Innovative Research and Engineering (INQUIRE) Lab, Norman, 73019, Oklahoma, U.S.A.
[2]School of Electrical and Computer Engineering, University of Oklahoma, Norman, 73019, Oklahoma, U.S.A.

*Corresponding author(s). E-mail(s): bana@ou.edu;
Contributing authors: gourab.datta-1@ou.edu; s.sh@ou.edu; shi@ou.edu;

**Abstract**

Metal–insulator-transition (MIT) materials are promising for switchable electronics, neuromorphic hardware, and reconfigurable photonics, but verified examples remain scarce and mechanistically complex. We argue that generative AI should target mechanism-informed phase-transition hypotheses rather than stable crystals alone. A modular physics-adapted framework, combined with staged verification of phase competition, electronic contrast, pathway plausibility, and control accessibility, can guide credible discovery under label scarcity.

## 1 Introduction

Metal–insulator-transition (MIT) materials occupy a distinctive place in condensed-matter physics and functional materials research. Small changes in temperature, pressure, strain, carrier density, dimensionality, composition or electric field can produce large changes in transport, optical response, magnetic order and lattice symmetry. This sensitivity makes MIT compounds attractive for switchable electronics, neuromorphic devices, sensing, reconfigurable photonics and phase-change functionality [1, 2]. It also makes them difficult to translate into devices. A material can host a scientifically compelling MIT yet remain unsuitable for a given application if its transition temperature, hysteresis, switching stimulus, contrast, thin-film compatibility or cycling stability falls outside the required operating window. New MIT discovery is therefore not only a search for additional correlated compounds, but a search for switchable phase landscapes that satisfy both physical and device-level constraints. Figure 1 places the verified MIT literature in chronological context, highlighting how the field evolved from individual transition compounds to broader material families with distinct correlation, charge-order, lattice-coupled and charge-transfer mechanisms.

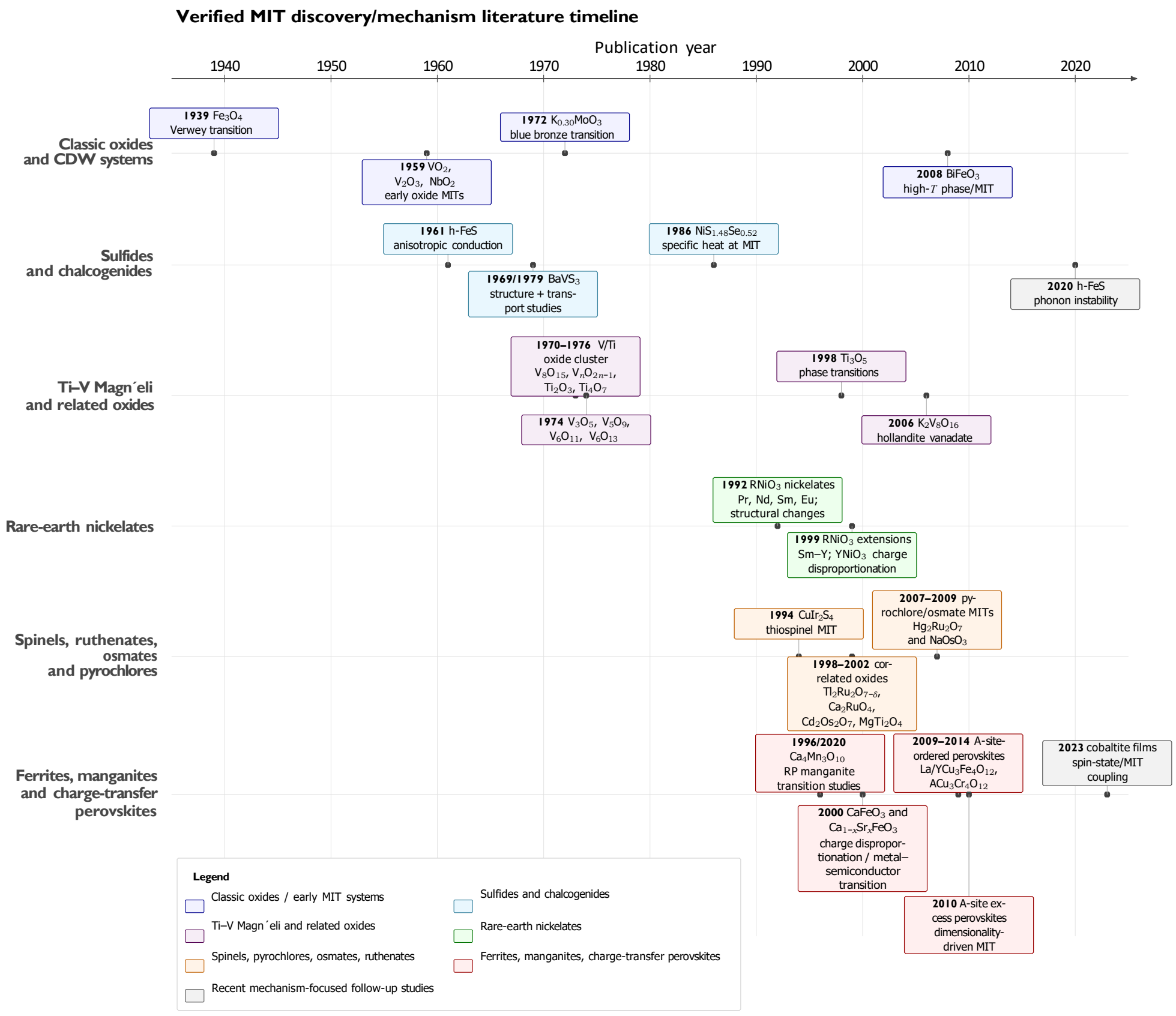

**Fig. 1 Historical timeline of representative metal–insulator transition literature [3–39].** The timeline summarizes early and systematic reports of MIT materials across oxide, chalcogenide, nickelate, cobaltite, and charge-transfer families.

The difficulty is physical as much as statistical. MITs do not arise from a single universal mechanism. In correlated materials, a transition may reflect competition between bandwidth and Coulomb repulsion; in other systems it can involve charge order, orbital order, Peierls-like distortion, spin-state changes, excitonic instability, disorder, dimensional confinement or electron–phonon coupling [40–42]. Even canonical examples rarely provide clean mechanism labels. Structural, electronic, magnetic and charge degrees of freedom can be inseparable, making MIT discovery different from ordinary property optimization: the target is not only a composition or a ground-state structure, but a switchable instability with an experimentally accessible and application-relevant control route.

The data landscape reflects this complexity. Large first-principles databases have transformed computational materials discovery by enabling high-throughput screening, property prediction and data-driven design across broad regions of inorganic chemistry [43]. Verified MIT labels, however, remain scarce. A curated database of thermally driven MIT-related compounds contains 343 materials labeled as metals, insulators or MIT compounds, with only 68 MIT materials at the time of publication [44]. Such labels require experimental evidence of a change in transport or optical character across a transition, often with structural, magnetic or spectroscopic interpretation. Their scarcity therefore reflects not only data availability, but also the cost, ambiguity and measurement dependence of establishing whether a material truly hosts a transition.

This places MIT discovery within inverse materials design, where the desired function is specified first and the task is to propose compositions and structures capable of realizing it [45, 46]. For MIT materials, the inverse-design target is more demanding than a static property such as formation energy, band gap or elastic modulus. The desired function is a controllable change in electronic state within a useful operating window. Thus, the target is not simply 'a small-gap material" or 'a correlated metal", but a material with at least two experimentally reachable phases of distinct electronic character, a plausible switching pathway and a control parameter compatible with the intended device context.

Generative artificial intelligence provides a route to search this space. Recent reviews of machine learning and generative modeling in materials science show that data-driven models can accelerate property prediction, inverse design and crystal-structure generation across broad chemical spaces [47, 48]. Crystal generative models can learn distributions of periodic structures and generate candidates beyond known compounds [49, 50]. Foundation-style inorganic materials models further show that diffusion generators can be pretrained across broad chemical spaces and adapted toward chemical, structural, electronic or magnetic constraints [51]. Yet these advances do not directly solve the MIT problem. Existing conditional generators are strongest when the target is a property of one structure: stability, composition, symmetry, band gap or magnetization [49–51]. MIT behavior is instead a relation between states. A generated compound with a small gap, transition-metal chemistry or low formation energy may be interesting, but none of these features demonstrates switchability, device-accessible control or reliable electronic contrast.

This distinction defines the gap addressed here. Prior machine-learning studies of MIT materials have curated datasets, identified descriptors and screened candidate oxides [44, 52, 53]. Modern crystal generators and large-scale stability models have expanded the space of plausible inorganic structures [49–51, 54–56]. Active-learning and autonomous-laboratory workflows have shown how closed-loop computation and experiment can accelerate materials discovery [57–60]. What remains missing is a generative framework in which the proposed object is not merely a stable crystal or high-scoring candidate, but a mechanism-informed phase-transition hypothesis with an explicit transition route, electronic contrast and device-relevant operating constraint.

In this Perspective, we argue for physics-adapted generative AI for MIT discovery. The core idea is modularity: a pretrained crystal generator supplies broad inorganic structural knowledge; a domain adaptation layer biases generation toward transition-metal oxides and chalcogenides; and mechanism-prior adapters steer the model using scalable physics proxies associated with correlation-driven, lattice-driven, charge-order, orbital-order, excitonic or mixed transition routes. Parameter-efficient adaptation methods such as low-rank adaptation offer a practical implementation without retraining the full generator for every physical route [61]. The aim is not to replace physical reasoning with a black-box model, but to encode physical hypotheses and device-relevant constraints as controllable generative biases.

Generated structures should therefore be treated first as transition hypotheses rather than as confirmed MIT discoveries. Claims of switchability require verification of thermodynamic accessibility, phase competition, electronic contrast, pathway plausibility and experimental control. Claims of device relevance require an additional check: whether the transition window, switching stimulus, hysteresis, stability and integration risks are compatible with a realistic use case. This Perspective first examines why MIT discovery is a scarce-label, multiphysics and device-constrained problem, then frames generation as inverse design of switchable phase landscapes, outlines a modular physics-adapted strategy, and proposes a verification-centered roadmap for converting generated structures into credible MIT candidates.

# 2 Why MIT discovery is a scarce-label, multiphysics and device-constrained problem

A central difficulty in metal–insulator-transition (MIT) discovery is that the label "MIT material" compresses several physical claims into one category. In many supervised materials-learning tasks, the target is a scalar property attached to a composition or one relaxed crystal structure. An MIT label is different: it implies at least two electronically distinct states, a control parameter that can connect them, and an experimentally measurable change in transport or optical response. For device-oriented discovery, the claim is even narrower: the transition should occur in a useful operating window, with sufficient contrast, reproducible switching and a control route compatible with integration. The relevant object is therefore not only a compound, but a compound together with its phase landscape, transition pathway, measurement context and intended operating constraint.

This definition explains why verified labels are scarce. A material cannot be classified confidently as an MIT compound simply because it has a small band gap, a high density of states near the Fermi level or a composition similar to known correlated oxides. It must show evidence of an electronically meaningful transition, usually through temperature, pressure, strain, field, doping or dimensionality dependent measurements. Even then, the mechanism may remain contested. $VO_2$ is commonly discussed in terms of coupled lattice distortion and electronic correlation, whereas rare-earth nickelates involve charge disproportionation, magnetism, octahedral distortions and negative charge-transfer physics [62, 63]. Layered chalcogenides further caution against clean labels: in 1T-$TaS_2$, charge-density-wave order, Mott localization, stacking and band-structure effects are intertwined [64, 65]; in $Ta_2NiSe_5$, the proposed excitonic instability is strongly coupled to a structural transition and the relative roles of electronic and lattice order remain debated [66–68]. These systems are therefore not pure templates for single mechanisms or universal device platforms, but examples of why MIT labels are composite and context-dependent.

This complexity has direct consequences for artificial intelligence. Confirmed positives are few: a curated database of thermally driven MIT-related materials contains hundreds of labeled examples, with only a minority assigned to the MIT class [44]. Such data are valuable for benchmarking and calibration, but too small to train a modern generative model directly on the MIT class. Negative labels are also uncertain, because a compound reported as metallic or insulating under one condition may transition under another pressure range, strain state, stoichiometry, dimensionality or doping level. In addition, the same composition may support several polymorphs, magnetic states or ordered configurations, only some of which are relevant to the transition. Device relevance adds another layer of sparsity: even a verified MIT may be unsuitable if its transition temperature, hysteresis, switching stimulus, thin-film stability or cycling response does not match the intended use. MIT discovery is therefore a poor fit for a binary generation objective.

Figure 2 summarizes this mismatch. General inorganic datasets provide abundant information about structural chemistry, whereas verified transition behavior is sparse. At the same time, the target becomes richer as one moves from stable-crystal generation to transition-candidate generation. A useful model must bridge large structural datasets and scarce transition labels, while also moving from broad chemical validity to narrow mechanism-prior instability and, ultimately, application-specific switching constraints. Scaling generic crystal data can improve crystallographic realism, but it does not by itself teach which electronic, lattice, magnetic, charge or orbital instability is needed for an MIT, or whether that instability can be accessed under realistic device conditions.

The diversity of known systems makes this point concrete. Figure 3 places representative MIT systems on a transition-temperature axis and highlights the different physical routes that can produce electronic switching. Full details of the 68 MIT-positive compounds are summarized in Supplementary Figure S1 and Supplementary Table S2. Table 1 compares representative oxides, chalcogenides, thiospinels, layered compounds and correlated perovskites. The purpose is not to assign each material a settled mechanism, but to show that structural distortion, electronic correlation, charge order, orbital polarization, dimensionality, magnetism and sample-dependent effects often coexist. These examples also show why known MIT compounds do not eliminate the need for discovery: many are excellent physical benchmarks, but their transition scales, switching routes or sample sensitivities may not match a target device. Two lessons follow. First, the relevant thermal energy scale is often tens of meV or less, although high-temperature systems such as $NbO_2$ show that larger scales can be relevant when a strong structural pathway exists. A generated phase pair separated by hundreds of meV per formula unit is therefore unlikely to be a credible room-temperature or low-power MIT hypothesis unless strain, pressure, entropy or kinetic trapping provides a specific route. Second, mechanism labels are not clean classes. $VO_2$, $V_2O_3$, nickelates, 1T-$TaS_2$ and $Ta_2NiSe_5$ show that structural, electronic, magnetic and lattice degrees of freedom can be inseparable. This supports mechanism-prior adapters trained from scalable proxies, not supervised adapters trained by splitting a small MIT dataset into rigid mechanism labels.

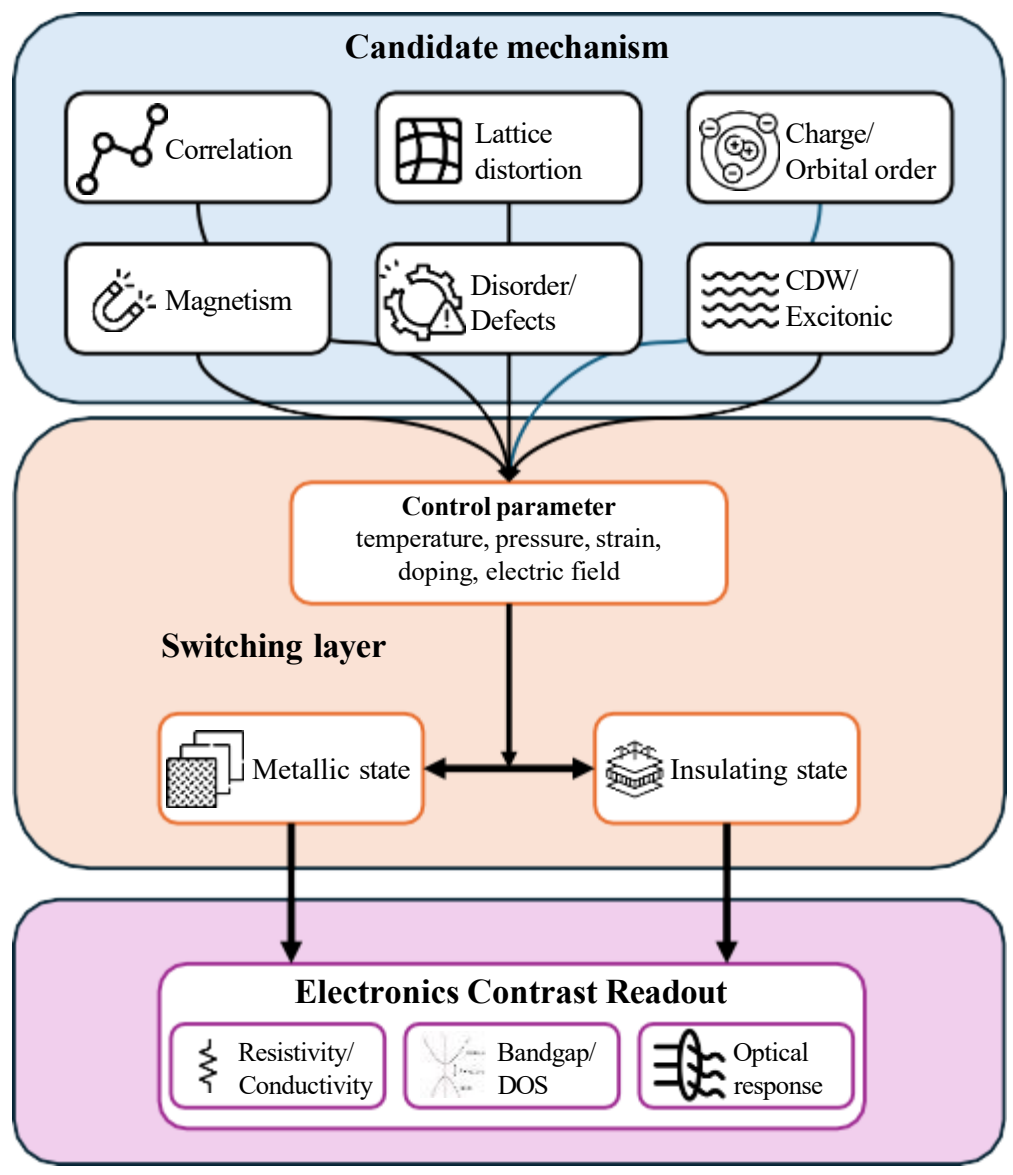

**Fig. 2 Metal–insulator-transition discovery as a scarce-label, multiphysics generative problem.** General materials databases provide large numbers of stable or metastable inorganic structures, whereas experimentally verified MIT labels are much rarer. At the same time, the discovery target becomes progressively more demanding: a valid crystal must also be thermodynamically plausible, support competing electronic or structural states, show measurable electronic contrast, and switch under experimentally accessible conditions. This creates a mismatch between the data distribution available to train generic crystal generators and the mechanism-rich subspace relevant to MIT discovery.

A more useful formulation is to treat MIT discovery as a conditional search over physical routes and operating constraints. The label can be decomposed into mechanism-relevant attributes such as correlation strength, bandwidth sensitivity, metal–ligand covalency, charge-transfer tendency, lattice dimerization, orbital degeneracy, dimensionality, charge ordering, magnetic ordering and disorder tolerance. None of these descriptors alone defines an MIT material, but together they describe the instabilities a generator should be encouraged to explore. The inverse-design target is therefore a constrained phase landscape rather than an isolated point in property space: a useful candidate must be structurally plausible, thermodynamically accessible, close to a competing phase, capable of electronic reconstruction and switchable by a realistic control parameter. For device-facing discovery, it must also offer a plausible transition window, contrast and integration pathway. This motivates a modular strategy in which a broad generator supplies chemical and crystallographic priors, while smaller adaptation modules specialize that prior toward relevant chemical families, mechanism-prior motifs and use-case constraints.

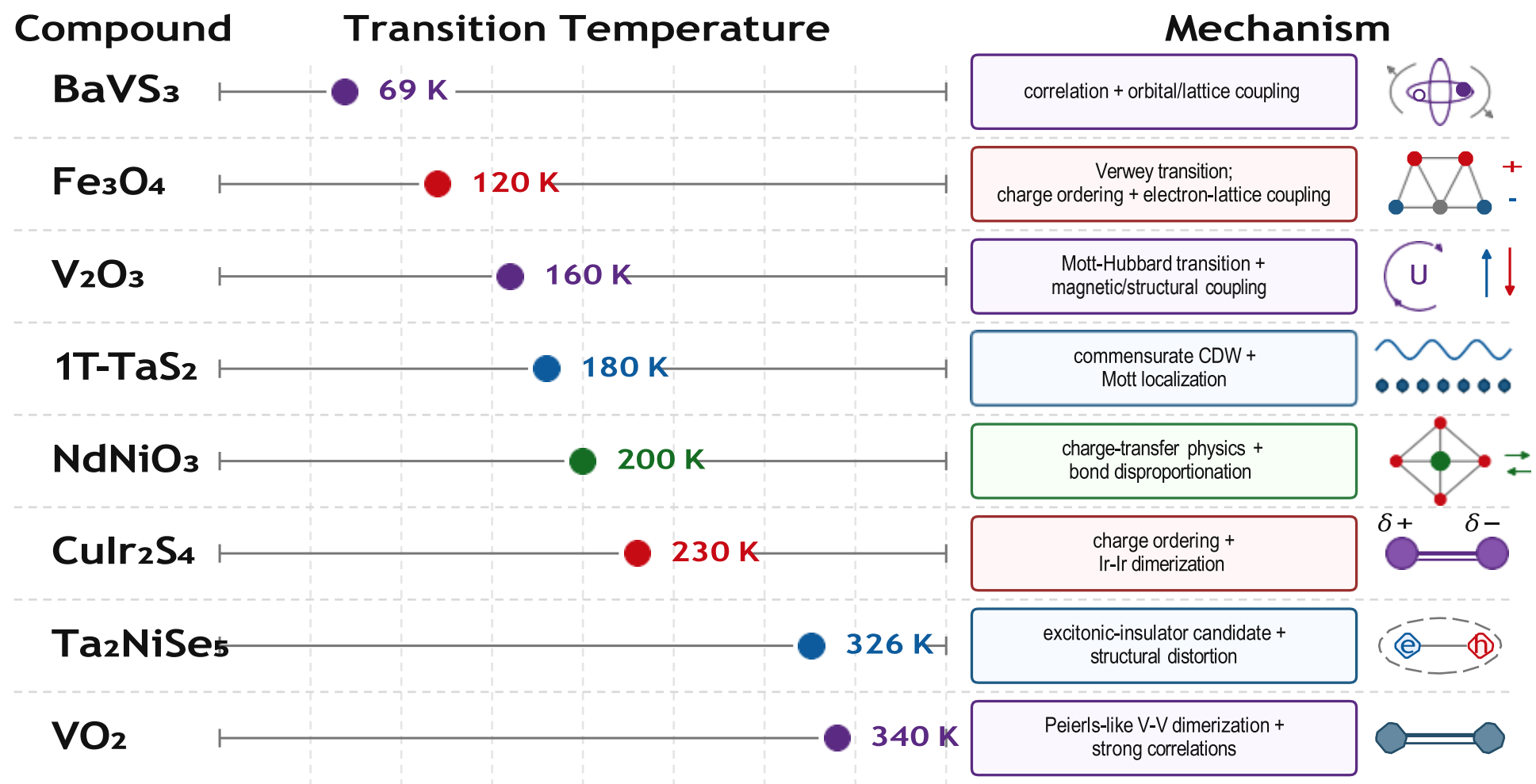

**Fig. 3 Representative MIT systems span broad thermal scales and coupled mechanisms.** Approximate transition temperatures are shown for selected MIT materials, together with the dominant or commonly discussed transition physics. The examples include correlation and orbital/lattice coupling in $BaVS_3$, Verwey charge ordering in $Fe_3O_4$, Mott–Hubbard physics in $V_2O_3$, charge-density-wave/Mott coupling in 1T-$TaS_2$, charge-transfer and bond disproportionation in $NdNiO_3$, charge ordering and Ir–Ir dimerization in $CuIr_2S_4$, excitonic–structural physics in $Ta_2NiSe_5$, and Peierls-like V–V dimerization with correlations in $VO_2$. The diversity reinforces that MIT generation should target mechanism-informed phase-transition hypotheses rather than a single universal label.

# 3 Physics-adapted generative AI for MIT materials

Generative artificial intelligence can reformulate metal–insulator-transition (MIT) discovery as inverse design of switchable phase landscapes. For many materials-design tasks, conditioning is attached to a static property, such as formation energy, band gap, elastic modulus or magnetic moment. MIT materials require a different target: a candidate whose phase landscape contains electronically distinct states connected by an experimentally accessible control parameter. For device-facing discovery, this target is narrower still, because the transition should also occur within a useful operating window and provide sufficient contrast, hysteresis control and integration compatibility. The generated object is therefore not only a crystal, but a transition hypothesis.

This distinction defines the contribution of a physics-adapted approach. Generic crystal generators can produce valid, novel and stable inorganic structures, but those metrics do not establish MIT relevance. A small band gap does not imply switchability, and transition-metal chemistry does not guarantee phase competition. The proposed shift is not the isolated use of a pretrained generator, domain fine-tuning, low-rank adaptation or density-functional-theory screening; each already has precedent. The shift is to organize these components around a proposed transition route, then verify whether the generated structure supports the corresponding phase competition, electronic contrast and realistic control pathway.

**Table 1 Representative MIT systems as phase-transition targets.** Approximate scales are reported as transition temperature and $k_B T$ where available. Controls are abbreviated as temperature ($T$), pressure ($P$), doping/substitution ($x$), strain ($\epsilon$), electric field/voltage ($E$), light ($h\nu$), thickness ($t$), and current ($I$). Mechanisms are kept coupled or contested where a single assignment is insufficient.

| System/form | Control/scale | Mechanism status and verification implication |
|---|---|---|
| $Fe_3O_4$; Verwey spinel [69] | $T$; 120–122 K / $\sim$ 10 meV. | Charge/orbital order, lattice distortion, and stoichiometry sensitivity; check charge/bond order, defects, and stoichiometry. |
| $VO_2$; canonical oxide MIT [4] | $T, \epsilon, E$; $\sim$ 340 K / $\sim$ 29 meV. | Rutile–monoclinic distortion, V–V dimerization, and correlations; verify metallic rutile-like and insulating monoclinic-like phases. |
| $V_2O_3$; Mott oxide [70] | $T, P, x, \epsilon$; 150–160 K / $\sim$ 13 meV. | Mott physics coupled to antiferromagnetism and monoclinic distortion; test magnetism, $U$, functional choice, and distortion jointly. |
| $R$NiO$_3$; rare-earth nickelates [63, 71] | $T, \epsilon, R, E$; family dependent. | Bond disproportionation, negative charge transfer, octahedral rotations, and magnetism; include strain and rare-earth dependence. |
| $NbO_2$; high-$T$ $d^1$ oxide [72] | $T, E$; $\sim$ 1080 K / $\sim$ 93 meV. | Peierls-like dimerization with electronic coupling; high transition scale is possible, but usefulness depends on the control route. |
| $BaVS_3$; quasi-1D sulfide [73, 74] | $T, P$; 69–70 K / $\sim$ 6 meV. | Structural instability, quasi-1D physics, and correlations; account for pressure, sample quality, and low energy scale. |
| 1T-$TaS_2$; layered CDW dichalcogenide [64, 65] | $T, h\nu, E, t$; $\lesssim$ 180 K / $\sim$ 16 meV. | CDW order intertwined with Mottness, stacking, and band effects; use as an ambiguity benchmark, not a pure Mott template. |
| $Ta_2NiSe_5$; excitonic candidate [66–68] | $T, t, E$; $\sim$ 328 K / $\sim$ 28 meV. | Excitonic instability versus structural/lattice origin remains debated; separate excitonic and lattice priors and require pathway validation. |
| $CuIr_2S_4$; thiospinel MIT [75, 76] | $T, x$; $\sim$ 230 K / $\sim$ 20 meV. | Charge order, Ir dimerization, and spin-singlet formation; target charge-order and dimerization priors. |
| $Ca_2RuO_4$; layered ruthenate [77, 78] | $T, P, I$; $\sim$ 357 K / $\sim$ 31 meV. | Structurally driven Mott transition with orbital–lattice coupling; check structural collapse, orbital polarization, and response. |
| Manganites; $La_{1-x}Ca_xMnO_3$ and related CMR systems [79, 80] | $x, T, E$; composition dependent. | Double exchange, Jahn–Teller coupling, phase competition, and percolation; specify control parameter and sample state. |

Figure 4 summarizes this design philosophy. Conditional generation is already becoming standard in materials design: crystal generators have been conditioned on composition, pressure, symmetry, local environment, stability and target properties, while large-scale graph-network models have expanded the catalogue of predicted stable crystals [54–56]. These advances are essential, but they do not define the MIT target. For MIT discovery, the conditioning signal must be tied to a candidate phase pair, a possible mechanism, a measurable electronic contrast and, when relevant, the operating constraint that makes the transition useful.

**Fig. 4 Physics-adapted inverse design for metal–insulator-transition materials. a)** Conventional screening starts from known or enumerated compounds and asks whether any exhibit MIT behavior. Generic crystal generation expands the pool of valid and stable structures, but does not establish switchability. MIT inverse design instead starts from the desired function: electronically distinct states connected by an accessible control parameter. **b)** A modular framework combines a pretrained crystal diffusion generator with lightweight adapters. The base model provides broad inorganic structural knowledge; a domain adapter biases generation toward transition-metal oxides and chalcogenides; and mechanism-prior adapters steer sampling using proxies for correlation, dimerization, charge or orbital order, excitonic or density-wave instability and mixed Mott–Peierls routes.

Table 2 makes the role of mechanism-prior adapters explicit. The equations are not used as hard labels or universal transition criteria; rather, they define physical coordinates along which a generator can be biased and a verifier can test candidates. This is important because an MIT candidate may satisfy more than one row. For example, a lattice distortion may narrow a band, a charge-density wave may coexist with correlation effects, and an excitonic interpretation may compete with a structural one. The useful target is therefore not a pure mechanism class, but a phase pair whose electronic contrast is supported by physically interpretable evidence and whose switching route can be assessed against the intended use case.

**Table 2 Physics priors for mechanism-aware MIT generation.** The relations are not universal MIT equations. They are compact physical handles that can be translated into descriptors, proxy labels, or verification tests for physics-adapted generation.

| Route | Minimal physical handle | Use in generation and verification |
|---|---|---|
| **Mott–Hubbard** | $U/W \gtrsim (U/W)_c$; Hubbard-like local repulsion. | Prioritizes narrow $d$ bands, localized orbitals and weak hopping; verify gap opening or suppressed $N(E_F)$ under reasonable $U$, magnetic and functional choices. |
| **Peierls/dimerization** | $q \simeq 2k_F$; $\Delta_P \propto gu$. | Prioritizes chain-like motifs and short metal–metal bonds; verify an undistorted/dimerized phase pair with electronic contrast and a soft distortion mode. |
| **Charge-density wave** | Large $\chi(q)$, often with finite-$q$ phonon softening. | Prioritizes layered or low-dimensional structures, nesting tendency and commensurate supercells; verify CDW-like distortion, symmetry lowering and electronic contrast. |
| **Charge order / bond disproportionation** | $H_V = V \sum_{\langle ij \rangle} n_i n_j$; $\Delta q, \Delta d \neq 0$. | Prioritizes mixed valence, inequivalent metal sites and breathing distortions; verify Bader/valence contrast, bond disproportionation and insulating behavior. |
| **Orbital / Jahn–Teller order** | $P = (n_a - n_b)/(n_a + n_b)$; $E_{JT} \sim g^2/2K$. | Prioritizes partially filled degenerate orbitals and cooperative distortions; verify orbital polarization, symmetry lowering and electronic reconstruction. |
| **Excitonic / semimetal instability** | $E_b \gtrsim E_g$; $\Delta_X = \langle c^\dagger v \rangle$. | Prioritizes small-gap or semimetallic parents with weak screening and distinct band-edge orbital characters; verify whether electronic instability is separable from structural gap opening. |
| **Disorder / percolative localization** | Ioffe–Regel limit $k_F \ell \sim 1$; short localization length $\xi$. | Prioritizes disorder tolerance, mixed occupancy, strong scattering and phase coexistence; verify transport contrast controlled by disorder, stoichiometry, microstructure or percolation. |

A practical implementation separates broad chemical realism from transition-specific steering. A pretrained crystal diffusion model supplies structural knowledge from large materials datasets, including coordination environments, lattice motifs, chemical compatibilities and periodic geometries [49–51]. This prior is essential because verified MIT compounds are too few to train a reliable generator directly on the positive class. Domain adaptation can then bias sampling toward transition-metal oxides and chalcogenides, where partially filled *d* orbitals, variable oxidation states, metal–ligand hybridization, lattice distortions and reduced dimensionality often support competing electronic phases. This restriction is a search prior, not a claim that MIT behavior is exclusive to these families.

The next layer is mechanism-prior adaptation. These adapters should not be trained by partitioning the small set of verified MIT compounds into mechanism classes. Such a strategy would be statistically fragile and physically misleading, because known MIT materials are few and many transitions involve intertwined mechanisms. Instead, mechanism-prior adapters should be trained or conditioned using scalable physics proxies computed over larger pools of candidate oxides and chalcogenides. Bandwidth and metal–ligand geometry can guide correlation-prior steering; metal–metal distance and symmetry-lowering distortions can guide Peierls-like steering; oxidation-state flexibility and bond disproportionation can guide charge-order steering; orbital degeneracy and crystal-field sensitivity can guide orbital-order steering; and small-gap, semimetallic or low-dimensional electronic structure can guide excitonic or density-wave steering. A mechanism-prior adapter is therefore not a supervised classifier of known MIT mechanisms, but a controllable bias toward a testable physical scenario.

Parameter-efficient adaptation methods such as low-rank adaptation provide a practical implementation of this modular idea [61]. The pretrained generator can remain fixed while lightweight adapters steer generation toward selected domains or mechanism-prior directions. The verified MIT set should define the target phenomenon, calibrate proxy choices and evaluate enrichment, rather than serve as the sole training source. Proxy labels may come from high-throughput structural descriptors and selected electronic-structure calculations, including oxidation-state assignments, density-of-states features, bond-disproportionation metrics, phonon soft modes and dimensionality descriptors. Device-oriented proxy labels could further encode transition-temperature range, likely control stimulus, dimensionality, strain sensitivity or thin-film relevance when such information is available.

The output of such a generator is not a discovered MIT material, but a structured hypothesis: a candidate crystal, a proposed mechanism-prior route, a possible competing phase, a suggested control parameter and an anticipated operating constraint. The generator improves the proposal distribution presented to the verification pipeline; physics-based computation and experiment determine whether any proposal is genuinely switchable and practically relevant. Key failure modes include collapse toward familiar chemistries, regeneration of structures close to the training set, biased proxy labels, unrealistic transition conditions and competing phases that arise from relaxation settings, magnetic initialization, Hubbard-$U$ choice or imposed symmetry breaking. Success should be measured as enrichment of credible transition hypotheses, not as direct discovery.

# 4 Verification before discovery

A physics-adapted generator can propose candidates, but it cannot establish discovery. This distinction is essential for metal–insulator-transition (MIT) materials, where a plausible composition or stable-looking structure is only the first filter. A generated crystal should instead be treated as a hypothesis that must pass a staged verification ladder: chemical validity, thermodynamic accessibility, phase competition, electronic contrast, pathway plausibility and experimental and device accessibility.

Figure 5 summarizes the proposed ladder and illustrates how its emphasis shifts across representative routes, including lattice-coupled, charge-density-wave/Mott-like, contested excitonic–structural and charge-order/dimerization-driven transitions. The ladder is not a rigid one-size-fits-all checklist. The same stages apply across systems, but the decisive evidence differs depending on whether the relevant instability is dominated by lattice distortion, charge ordering, low-dimensional symmetry breaking or strongly contested electronic structure. These examples also clarify why mechanism-prior adapters should be treated as overlapping generative biases rather than as hard labels. In each case, what matters is not whether a candidate matches a textbook category, but whether it supports an experimentally reachable phase pair with measurable electronic contrast and a plausible control route.

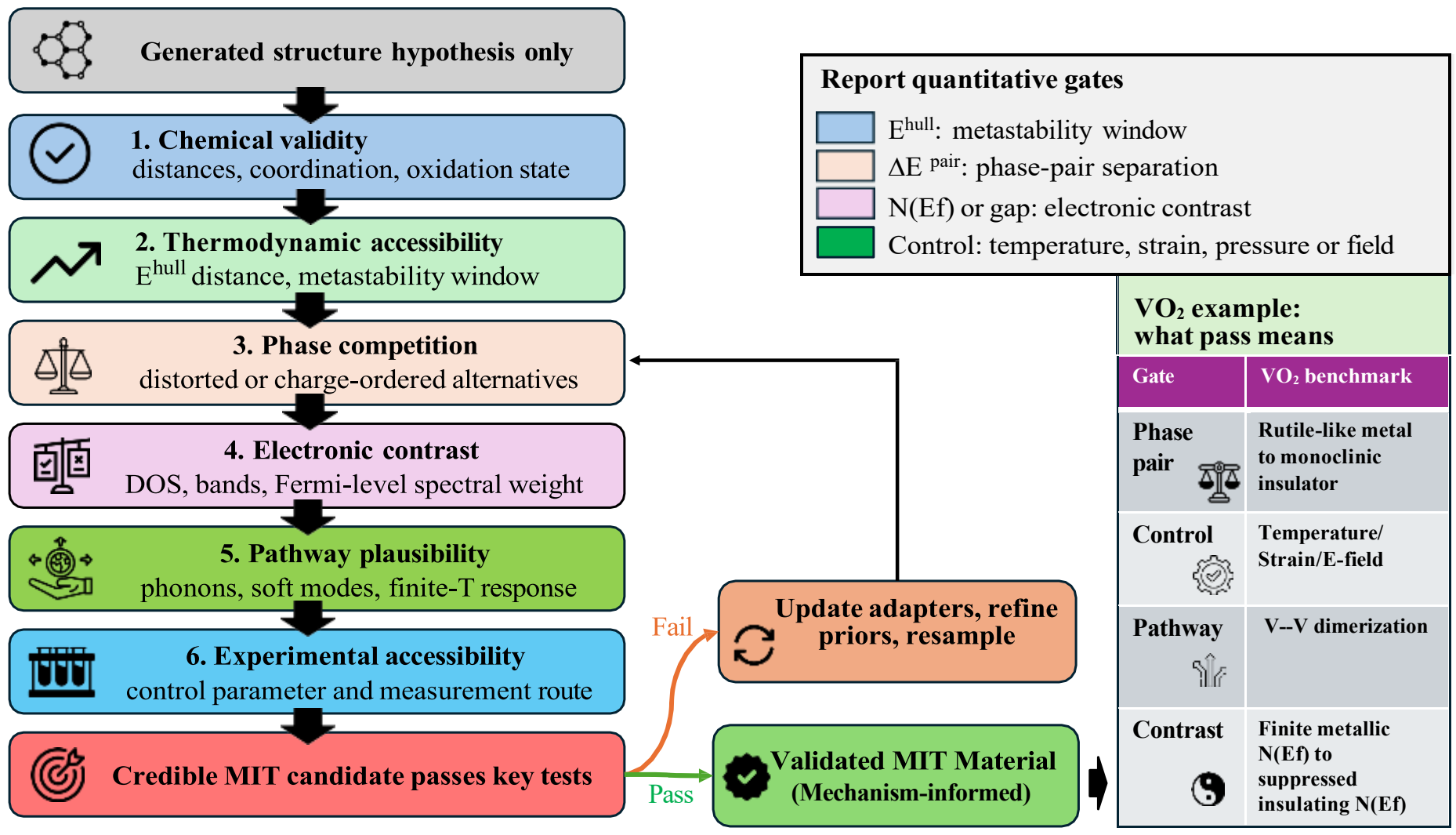

**Fig. 5 Verification ladder for generated MIT candidates.** Generated structures should move through increasingly stringent tests before being described as MIT candidates. Early filters assess structural validity and thermodynamic accessibility. Intermediate tests examine phase competition, symmetry-lowering distortions, magnetic or charge order and dynamical stability. Later stages test electronic contrast, transition pathway and experimental accessibility. In $VO_2$, this corresponds to asking whether metallic rutile-like and insulating monoclinic-like phases are structurally related, close enough in free energy to be accessible, electronically distinct and connected by a plausible distortion coordinate.

$VO_2$ illustrates what the ladder demands. The target is not simply 'generate $VO_2$" or 'generate a small-gap vanadium oxide", but a phase-pair hypothesis: a high-temperature rutile-like metallic phase and a low-temperature monoclinic insulating phase connected by V–V dimerization and lattice distortion near the observed transition regime [62]. A generated $VO_2$-like candidate would need chemically sensible, charge-balanced and structurally related phases, an accessible energy scale rather than a competing phase hundreds of meV per formula unit away, and electronic contrast between finite Fermi-level spectral weight and an opened or strongly suppressed insulating density of states. For device relevance, the same candidate would also need an operating window, hysteresis and switching route compatible with the intended use case. The example captures the general requirement: credible MIT hypotheses need a phase pair, an energy scale, electronic contrast, a switching coordinate and an experimentally meaningful control condition.

The first stage is structural and chemical validity: sensible interatomic distances, oxidation-state plausibility, reasonable coordination and absence of obvious crystallographic artifacts. The second stage is thermodynamic and metastability assessment. Energy above the convex hull, decomposition products and sensitivity to chemical potential should be reported quantitatively, with synthesis route, epitaxy and kinetic trapping treated as explicit justifications rather than assumed exceptions.

The third stage, phase competition, is where MIT verification departs from ordinary stable-material screening. A single relaxed structure with a small band gap or high density of states is insufficient. The candidate should support a plausible competing state: a distorted phase, magnetic variant, charge-ordered arrangement, orbital-ordered configuration, altered bond pattern or strain-/pressure-stabilized polymorph. For correlated transition-metal systems, this may require testing magnetic configurations, Hubbard-$U$ values, spin states and symmetry-broken initializations [81]. The fourth stage is electronic contrast. Candidate phases should show a meaningful change in electronic character, such as a metal-to-insulator, insulator-to-metal or semimetal-to-insulator transition. Density of states, band structure, orbital character and Fermi-level spectral weight provide early evidence, but robust candidates should preserve the qualitative contrast across reasonable computational choices.

Operational thresholds should be reported rather than left implicit. As early triage, candidates within $\sim 50$ meV atom$^{-1}$ of the convex hull can be treated as strong metastability candidates, whereas candidates above $\sim 100$–$200$ meV atom$^{-1}$ require a specific non-equilibrium, epitaxial or kinetic-stabilization route. Accessible phase pairs should preferably be separated by $\sim 25$–$50$ meV per formula unit or active transition-metal site, with larger separations requiring strain, pressure, entropy or magnetic-order justification. Electronic contrast should be reported as a gap opening, at least an order-of-magnitude suppression of $N(E_F)$, or a conductivity/resistivity contrast when transport proxies are available. These values are not universal definitions of an MIT; they are reporting gates for early-stage screening.

The fifth stage is pathway validation. If the mechanism involves lattice distortion, dimerization, charge-density-wave formation or symmetry breaking, phonon calculations and mode analysis can identify unstable or soft modes connecting candidate phases [82]. Thermal transitions may require finite-temperature effects, anharmonicity, entropy or molecular dynamics; strain-, pressure- or field-driven transitions should be evaluated under the corresponding constraint. The final stage is experimental and device accessibility. A useful candidate should include the proposed transition driver, approximate energy scale, likely transition direction, predicted electronic contrast and measurement route. Device-facing claims should additionally state the expected operating window, hysteresis requirement, likely switching stimulus, thin-film or interface sensitivity, thermal-management risk and cycling or variability concerns. Transport calculations can provide conductivity proxies when interpreted cautiously [83], but validation ultimately requires experimental evidence such as temperature-dependent resistivity, optical conductivity, structural characterization, spectroscopy or device-level switching.

The same ladder also evaluates generative models. Instead of reporting only validity, novelty and predicted stability, MIT generators should report enrichment across the ladder: the fraction of candidates passing chemical, stability, phase-competition, electronic-contrast, pathway and accessibility tests. The model's contribution is therefore measurable but modest: it improves the front end of discovery by increasing the fraction of generated candidates that survive MIT-specific verification, rather than declaring generated structures as discovered MIT materials.

# 5 Outlook: building a closed-loop ecosystem for MIT discovery

The next step for generative artificial intelligence in metal–insulator-transition (MIT) discovery is not simply to build larger models. Larger generators may improve structural realism and chemical coverage, but the central bottleneck remains that MIT behavior is rare, mechanism-dependent, device-constrained and difficult to verify. Progress will require a closed-loop ecosystem in which generative models, physics-based simulation, curated experimental data, targeted synthesis and device-level feedback inform one another.

Closed-loop materials discovery already provides a template. Active learning uses uncertainty and acquisition functions to select the next computation or experiment, while autonomous laboratories combine prediction, synthesis, characterization and feedback into iterative discovery cycles [57–60]. MIT materials make this loop more demanding because the objective is not a single property. Each iteration should update the proposed mechanism-prior route, candidate phase pair, control parameter, uncertainty in electronic contrast and compatibility with the intended operating window.

Three priorities follow. First, MIT datasets should move beyond binary labels. Useful records should include structure, polymorph, transition temperature, hysteresis, control parameter, measurement type, structural distortion, magnetic or charge order and the evidence used to assign the mechanism. For device-facing discovery, records should also capture film thickness, substrate, strain state, stoichiometry, switching stimulus, cycling response and measurement geometry when available. Uncertain and negative cases are equally important, because a material reported as metallic or insulating under one condition may transition under another pressure, strain, stoichiometry, dimensionality or doping range.

**Table 3 Mechanism-aware case studies for the verification ladder.** These examples are not intended as clean mechanism labels. Instead, they illustrate how the same verification ladder places different emphasis on phase competition, electronic contrast, and pathway evidence depending on the dominant or contested physics of a candidate system.

| System | Verification focus and implication for generative design |
|---|---|
| $VO_2$ [62] | *Lattice-coupled electronic transition; rutile–monoclinic phase competition.* The ladder should verify structurally related metallic rutile-like and insulating monoclinic-like phases, accessible phase-pair separation, a V–V dimerization pathway, and clear contrast between finite and suppressed $N(E_F)$. A generator should propose a phase pair and switching coordinate, not only a small-gap vanadium oxide. |
| 1T-$TaS_2$ [64, 65] | *Intertwined charge-density-wave, Mott-like, and stacking physics; mechanism contested.* The ladder should test for a symmetry-broken or CDW-like competing state, robust insulating character across stacking and magnetic choices, and electronic contrast that is not an artifact of one initialization. This case is an ambiguity benchmark rather than a pure Mott template. |
| $Ta_2NiSe_5$ [66–68] | *Contested excitonic–structural instability.* The ladder should test whether a low-gap or semimetallic parent develops an electronically distinct distorted state, whether the structural pathway is real, and whether an excitonic interpretation adds evidence beyond a lattice-driven transition. Mechanism-prior steering should treat excitonic and structural routes as overlapping priors requiring pathway validation. |
| $CuIr_2S_4$ [75, 76] | *Charge order, Ir dimerization, and spin-singlet formation in a thiospinel.* The ladder should test for charge disproportionation, dimerized Ir sublattices, insulating contrast relative to a higher-symmetry metallic state, and realistic phase-pair energetics near the thermal transition scale. This case shows why chalcogenide and thiospinel candidates require bond-order and charge-order verification beyond stability. |

Second, the field needs phase-pair benchmarks. Most materials databases are organized around single relaxed structures, whereas MIT discovery is about switching between states. Benchmarks should include metallic and insulating phases, high- and low-temperature polymorphs, symmetry-broken distortions, magnetic configurations, charge-ordered variants or strain-stabilized phases when available. Such data would test whether a generator proposes plausible electronic contrast between accessible states, rather than merely producing stable correlated crystals. Where possible, these benchmarks should also record the transition scale and control route, because the same phase pair may be useful for one device context but impractical for another.

Third, MIT generative studies should adopt standardized reporting. Beyond validity, novelty and predicted stability, they should report how many candidates pass each stage of the verification ladder: chemical plausibility, thermodynamic or metastable accessibility, phase competition, electronic contrast, pathway plausibility and experimental and device accessibility. This would make generative strategies comparable and reveal where a model fails: chemistry, stability, mechanism-prior steering, phase competition, electronic reconstruction, transition pathway or measurement feasibility.

The broader opportunity is to change what artificial intelligence is asked to do in quantum-material discovery. The goal should not be to replace physical intuition or declare discovery from generated structures alone. Rather, generative AI should formulate and prioritize physically testable hypotheses. For MIT materials, this means designing candidate phase landscapes in which electronic, structural, magnetic, orbital or charge degrees of freedom are close enough in energy to be switched by realistic external control, while also satisfying the operating constraints of a target application. Used this way, artificial intelligence becomes a mechanism-aware proposal engine for exploring rare regions of switchable electronic matter and identifying which candidates merit the next computation, synthesis or device test.

# Acknowledgements

This material is based upon work supported by the Air Force Office of Scientific Research under award number FA9550-25-1-0322. The funder had no role in the preparation of the manuscript, interpretation of the literature, or decision to submit the work for publication.

# Competing interests

The authors declare that they have no competing interests.

## Author contributions

Conceptualization, G.D. and Y.B.; methodology, G.D. and Y.B.; formal analysis, G.D., Z.S. and S.S.; investigation, G.D., Z.S. and S.S.; data curation, G.D.; writing, original draft preparation, G.D.; writing review and editing, Z.S., S.S. and Y.B.; visualization, G.D. and S.S.; supervision, Y.B., Z.S. and S.S.; project administration, Y.B., Z.S. and S.S. All authors have read and agreed to the published version of the manuscript.

## Data availability

Data sharing is not applicable to this Perspective because no new datasets were generated or analyzed. All literature sources discussed are cited in the manuscript and Supplementary Information.

## Code availability

Code availability is not applicable to this Perspective because no new custom code or computational workflow was developed for the manuscript.

# Supplementary file: Physics-adapted generative AI for metal–insulator transitions: designing switchable phases under label scarcity

Gourab Datta[1,2], Sarah Sharif[1,2], Zhisheng Shi[2], Yaser Banad[1,2*]

[1]Intelligent Neuromorphic and Quantum Understanding for Innovative Research and Engineering (INQUIRE) Lab, Norman, 73019, Oklahoma, U.S.A.

[2]School of Electrical and Computer Engineering, University of Oklahoma, Norman, 73019, Oklahoma, U.S.A.

*Corresponding author(s). E-mail(s): bana@ou.edu;
Contributing authors: gourab.datta-1@ou.edu; s.sh@ou.edu; shi@ou.edu;

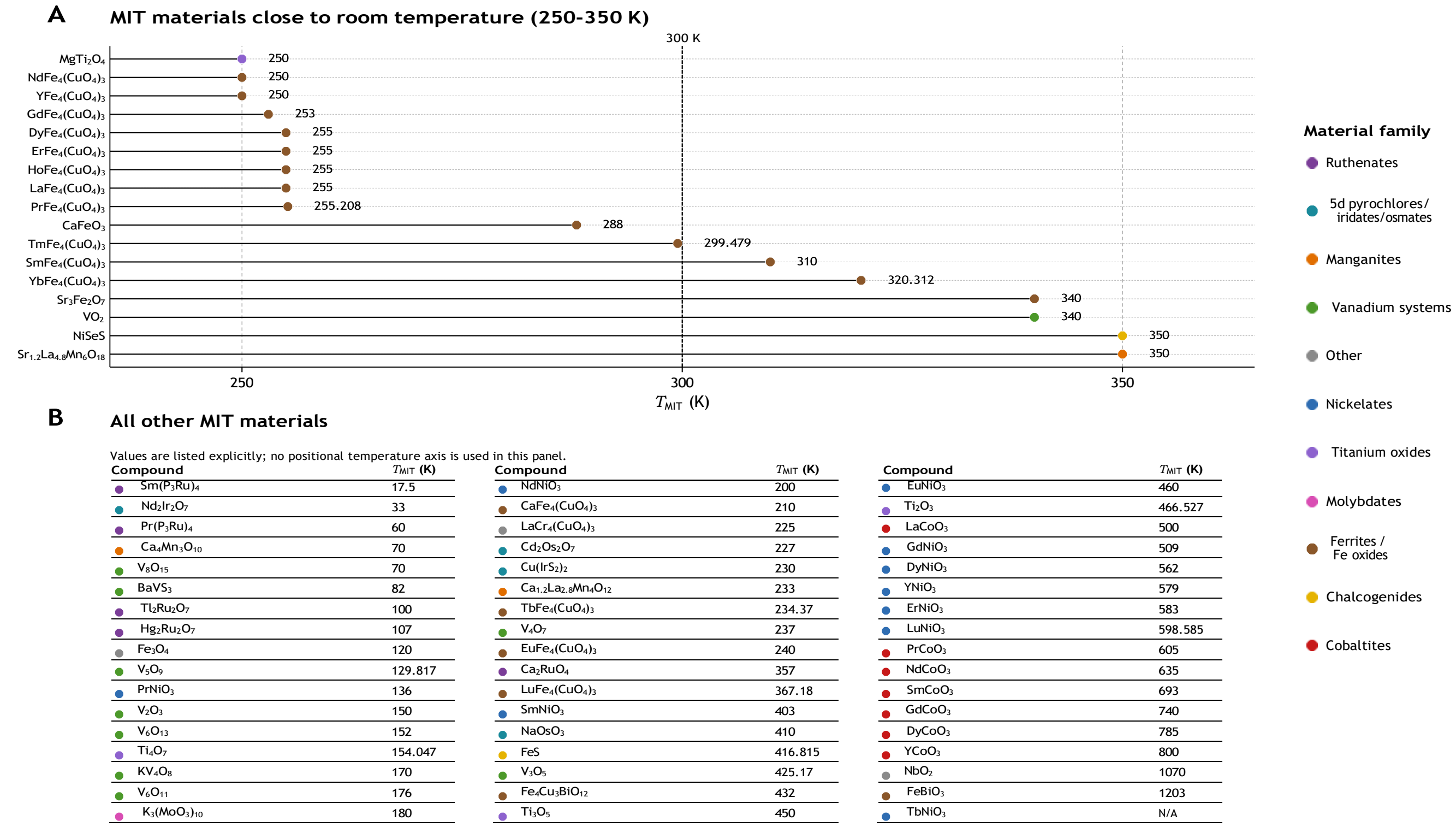

| Compound | $T_{MIT}$ (K) | Compound | $T_{MIT}$ (K) | Compound | $T_{MIT}$ (K) |
|---|---|---|---|---|---|
| $Sm(P_3Ru)_4$ | 17.5 | $NdNiO_3$ | 200 | $EuNiO_3$ | 460 |
| $Nd_2Ir_2O_7$ | 33 | $CaFe_4(CuO_4)_3$ | 210 | $Ti_2O_3$ | 466.527 |
| $Pr(P_3Ru)_4$ | 60 | $LaCr_4(CuO_4)_3$ | 225 | $LaCoO_3$ | 500 |
| $Ca_4Mn_3O_{10}$ | 70 | $Cd_2Os_2O_7$ | 227 | $GdNiO_3$ | 509 |
| $V_8O_{15}$ | 70 | $Cu(IrS_2)_2$ | 230 | $DyNiO_3$ | 562 |
| $BaVS_3$ | 82 | $Ca_{1.2}La_{2.8}Mn_4O_{12}$ | 233 | $YNiO_3$ | 579 |
| $Tl_2Ru_2O_7$ | 100 | $TbFe_4(CuO_4)_3$ | 234.37 | $ErNiO_3$ | 583 |
| $Hg_2Ru_2O_7$ | 107 | $V_4O_7$ | 237 | $LuNiO_3$ | 598.585 |
| $Fe_3O_4$ | 120 | $EuFe_4(CuO_4)_3$ | 240 | $PrCoO_3$ | 605 |
| $V_5O_9$ | 129.817 | $Ca_2RuO_4$ | 357 | $NdCoO_3$ | 635 |
| $PrNiO_3$ | 136 | $LuFe_4(CuO_4)_3$ | 367.18 | $SmCoO_3$ | 693 |
| $V_2O_3$ | 150 | $SmNiO_3$ | 403 | $GdCoO_3$ | 740 |
| $V_6O_{13}$ | 152 | $NaOsO_3$ | 410 | $DyCoO_3$ | 785 |
| $Ti_4O_7$ | 154.047 | $FeS$ | 416.815 | $YCoO_3$ | 800 |
| $KV_4O_8$ | 170 | $V_3O_5$ | 425.17 | $NbO_2$ | 1070 |
| $V_6O_{11}$ | 176 | $Fe_4Cu_3BiO_{12}$ | 432 | $FeBiO_3$ | 1203 |
| $K_3(MoO_3)_{10}$ | 180 | $Ti_3O_5$ | 450 | $TbNiO_3$ | N/A |

**Fig. 1 Supplementary Figure S1 — Transition-temperature grouping of MIT-positive compounds in the Georgescu et al. dataset.** The 68 metal–insulator transition (MIT)-positive compounds are separated according to proximity to room temperature. **a**, Compounds with reported transition temperatures in the near-room-temperature window, defined here as $250 \leq T_{\mathrm{MIT}} \leq 350$ K. Marker positions indicate the reported $T_{\mathrm{MIT}}$ values, and the dashed vertical line marks 300 K. **b**, Remaining MIT-positive compounds with reported transition temperatures outside the 250–350 K window, together with the entry for TbNiO3, for which no $T$ MIT value is reported in the processed summary. Panel b is shown as a categorical list with explicit temperature labels rather than a scaled temperature axis to avoid visual compression by high-temperature outliers. Colors denote broad material families. The figure is intended as a dataset-level overview of transition-temperature coverage and chemical diversity, not as a mechanism-resolved classification.

**Table 1**: **Supplementary Table S1 — Metal–insulator transition compounds, reported transition temperatures and representative mechanisms.** The 68 entries correspond to the MIT-positive subset of the Georgescu/Rondinelli processed MIT dataset used in this work. $T_{\text{MIT}}$ values are reported as listed in the processed summary file. Mechanism labels are intentionally written as dominant or representative assignments rather than mutually exclusive categories because many MITs involve coupled electronic, magnetic, charge, orbital and lattice degrees of freedom.

| No. | MIT compound | Dataset formula | $T_{\text{MIT}}$ (K) | Representative mechanism class | Dominant or reported transition physics | Reference key(s) |
|---|---|---|---|---|---|---|
| 1 | $SmRu_4P_{12}$ | $Sm(P_3Ru)_4$ | 17.5 | Filled skutterudite; correlated/multipolar order | Metal–insulator transition associated with f-electron/cage hybridization, orbital or multipolar ordering, and lattice/electronic reconstruction. | [1] |
| 2 | $Nd_2Ir_2O_7$ | $Nd_2Ir_2O_7$ | 33 | Spin–orbit-coupled Mott/Slater physics | Pyrochlore iridate transition commonly associated with all-in–all-out antiferromagnetism, spin–orbit coupling, and correlation-driven gap opening. | [1] |
| 3 | $PrRu_4P_{12}$ | $Pr(P_3Ru)_4$ | 60 | Filled skutterudite; charge/orbital ordering | Skutterudite MIT generally linked to ordering of electronic degrees of freedom coupled to lattice distortion and hybridization. | [1] |
| 4 | $Ca_4Mn_3O_{10}$ | $Ca_4Mn_3O_{10}$ | 70 | Manganite; magnetic/electron–lattice coupling | Ruddlesden–Popper manganite transition with magnetoresistance switching and glassy magnetic/electron–lattice coupling. | [2, 3] |
| 5 | $V_8O_{15}$ | $V_8O_{15}$ | 70 | Magn´eli vanadium oxide; charge localization | Mixed-valence V oxide MIT with charge localization/order and V–V bonding/lattice distortion. | [4, 5] |
| 6 | $BaVS_3$ | $BaVS_3$ | 82 | Quasi-1D correlated sulfide | Transition involving one-dimensional correlations, orbital/lattice coupling, and possible density-wave-like instability. | [6, 7] |
| 7 | $Tl_2Ru_2O_7$ | $Tl_2Ru_2O_7$ | 100 | Pyrochlore ruthenate; correlation/structural coupling | Metal–semiconductor transition in a pyrochlore ruthenate with Ru–O network reconstruction and correlation effects. | [8] |
| 8 | $Hg_2Ru_2O_7$ | $Hg_2Ru_2O_7$ | 107 | Pyrochlore ruthenate; correlation/structural coupling | Pyrochlore-type ruthenate MIT with coupled charge, lattice, and magnetic degrees of freedom. | [9] |

*Continued on next page*

**Table 1: Supplementary Table S1 — Continued.**

| No. | MIT compound | Dataset formula | $T_{MIT}$ (K) | Representative mechanism class | Dominant or reported transition physics | Reference key(s) |
|---|---|---|---|---|---|---|
| 9 | $Fe_3O_4$ | $Fe_3O_4$ | 120 | Verwey transition; charge/orbital order | Classic Verwey transition with $Fe^{2+}/Fe^{3+}$ charge/orbital ordering and strong electron–lattice coupling. | [10] |
| 10 | $V_5O_9$ | $V_5O_9$ | 129.8 | Magnéli vanadium oxide; Mott–Peierls/charge order | Mixed-valence transition with electron localization, V–V pairing, and lattice distortion. | [5, 11] |
| 11 | $PrNiO_3$ | $PrNiO_3$ | 136 | Rare-earth nickelate; charge-transfer/bond disproportionation | Insulator–metal transition governed by charge-transfer physics, Ni–O bond disproportionation, and antiferromagnetic order. | [12, 13] |
| 12 | $V_2O_3$ | $V_2O_3$ | 150 | Mott–Hubbard transition | Canonical correlated MIT involving Hubbard physics, magnetic ordering, and structural change. | [14] |
| 13 | $V_6O_{13}$ | $V_6O_{13}$ | 152 | Magnéli vanadium oxide; charge localization | Mixed-valence transition with charge localization/order, correlation, and lattice distortion. | [5, 11] |
| 14 | $Ti_4O_7$ | $Ti_4O_7$ | 154.0 | Titanium Magnéli phase; bipolaron/charge order | MIT associated with $Ti^{3+}$–$Ti^{3+}$ pairing, bipolaron formation, charge order, and structural distortion. | [15] |
| 15 | $K_2V_8O_{16}$ | $KV_4O_8$ | 170 | Hollandite vanadate; charge order/lattice distortion | Dataset stoichiometry $KV_4O_8$ corresponds to $K_2V_8O_{16}$; transition involves charge ordering and structural distortion. | [16] |
| 16 | $V_6O_{11}$ | $V_6O_{11}$ | 176 | Magnéli vanadium oxide; Mott–Peierls/charge order | Mixed-valence transition with electron localization, V–V pairing, and lattice distortion. | [5, 11] |
| 17 | $K_{0.30}MoO_3$ / $K_3(MoO_3)_{10}$ | $K_3(MoO_3)_{10}$ | 180 | Blue bronze; Peierls/CDW transition | Quasi-one-dimensional molybdenum bronze transition dominated by a charge-density-wave/Peierls instability; early literature also discussed excitonic-insulator character. | [17] |
| 18 | $NdNiO_3$ | $NdNiO_3$ | 200 | Rare-earth nickelate; charge-transfer/bond disproportionation | Charge-transfer nickelate MIT with Ni–O bond disproportionation and antiferromagnetic/electronic reconstruction. | [12, 13] |
| 19 | $CaCu_3Fe_4O_{12}$ | $CaFe_4(CuO_4)_3$ | 210 | A-site ordered perovskite; intersite charge transfer | Temperature-driven Cu–Fe intersite charge transfer / charge disproportionation coupled to lattice changes. | [18] |

*Continued on next page*

**Table 1**: **Supplementary Table S1 — Continued.**

| No. | MIT compound | Dataset formula | $T_{\mathrm{MIT}}$ (K) | Representative mechanism class | Dominant or reported transition physics | Reference key(s) |
|---|---|---|---|---|---|---|
| 20 | $LaCu_3Cr_4O_{12}$ | $LaCr_4(CuO_4)_3$ | 225 | A-site ordered chromate; intersite charge transfer | Temperature-induced intersite charge transfer involving Cr ions in A-site-ordered $ACu_3Cr_4O_{12}$. | [19] |
| 21 | $Cd_2Os_2O_7$ | $Cd_2Os_2O_7$ | 227 | 5d pyrochlore; Slater/Mott-like AFM transition | Continuous pyrochlore osmate MIT associated with magnetic ordering, spin–orbit coupling, and correlation effects. | [20] |
| 22 | $CuIr_2S_4$ | $Cu(IrS_2)_2$ | 230 | Thiospinel; charge order and dimerization | MIT associated with Ir charge ordering/orbital molecules and Ir–Ir dimerization in a thiospinel lattice. | [21] |
| 23 | $Ca_{1.2}La_{2.8}Mn_4O_{12}$ | $Ca_{1.2}La_{2.8}Mn_4O_{12}$ | 233 | A-site ordered manganite; dimensionality-driven | Insulator–metal transition in A-site excess/non-stoichiometric perovskite manganite linked to dimensionality, Mn valence, and spin/lattice coupling. | [22] |
| 24 | $TbCu_3Fe_4O_{12}$ | $TbFe_4(CuO_4)_3$ | 234.4 | A-site ordered ferrite; intersite charge transfer | Rare-earth Cu–Fe quadruple perovskite transition dominated by Cu–Fe intersite charge transfer and Fe charge disproportionation. | [18] |
| 25 | $V_4O_7$ | $V_4O_7$ | 237 | Magn´eli vanadium oxide; Mott–Peierls/charge order | Mixed-valence V oxide transition with charge localization, V–V pairing, and lattice reconstruction. | [5] |
| 26 | $EuCu_3Fe_4O_{12}$ | $EuFe_4(CuO_4)_3$ | 240 | A-site ordered ferrite; intersite charge transfer | Cu–Fe intersite charge transfer / Fe charge disproportionation with lattice coupling. | [18] |
| 27 | $MgTi_2O_4$ | $MgTi_2O_4$ | 250 | Spinel titanate; orbital-Peierls/dimerization | Metal-to-spin-singlet-insulator transition involving Ti–Ti dimerization, orbital ordering, and frustration relief on a pyrochlore sublattice. | [23] |
| 28 | $NdCu_3Fe_4O_{12}$ | $NdFe_4(CuO_4)_3$ | 250 | A-site ordered ferrite; intersite charge transfer | Rare-earth Cu–Fe quadruple perovskite with Cu–Fe charge transfer and Fe charge disproportionation. | [18] |
| 29 | $YCu_3Fe_4O_{12}$ | $YFe_4(CuO_4)_3$ | 250 | A-site ordered ferrite; charge disproportionation | Charge-disproportionated perovskite where intersite charge transfer is suppressed/modified relative to other $RCu_3Fe_4O_{12}$ members. | [18, 24] |
| 30 | $GdCu_3Fe_4O_{12}$ | $GdFe_4(CuO_4)_3$ | 253 | A-site ordered ferrite; intersite charge transfer | Cu–Fe intersite charge transfer / Fe charge disproportionation with lattice coupling. | [18] |
| 31 | $DyCu_3Fe_4O_{12}$ | $DyFe_4(CuO_4)_3$ | 255 | A-site ordered ferrite; intersite charge transfer | Cu–Fe intersite charge transfer / Fe charge disproportionation with lattice coupling. | [18] |

*Continued on next page*

**Table 1: Supplementary Table S1 — Continued.**

| No. | MIT compound | Dataset formula | $T_{MIT}$ (K) | Representative mechanism class | Dominant or reported transition physics | Reference key(s) |
|---|---|---|---|---|---|---|
| 32 | $ErCu_3Fe_4O_{12}$ | $ErFe_4(CuO_4)_3$ | 255 | A-site ordered ferrite; intersite charge transfer | Cu–Fe intersite charge transfer / Fe charge disproportionation with lattice coupling. | [18] |
| 33 | $HoCu_3Fe_4O_{12}$ | $HoFe_4(CuO_4)_3$ | 255 | A-site ordered ferrite; intersite charge transfer | Cu–Fe intersite charge transfer / Fe charge disproportionation with lattice coupling. | [18] |
| 34 | $LaCu_3Fe_4O_{12}$ | $LaFe_4(CuO_4)_3$ | 255 | A-site ordered ferrite; intersite charge transfer | Temperature-induced Cu–Fe intersite charge transfer in an A-site-ordered perovskite. | [18, 25] |
| 35 | $PrCu_3Fe_4O_{12}$ | $PrFe_4(CuO_4)_3$ | 255.2 | A-site ordered ferrite; intersite charge transfer | Cu–Fe intersite charge transfer / Fe charge disproportionation with lattice coupling. | [18] |
| 36 | $CaFeO_3$ | $CaFeO_3$ | 288 | Ferrate; charge disproportionation | Metal–semiconductor transition associated with Fe charge disproportionation, breathing distortion, and magnetic order. | [26, 27] |
| 37 | $TmCu_3Fe_4O_{12}$ | $TmFe_4(CuO_4)_3$ | 299.5 | A-site ordered ferrite; intersite charge transfer | Cu–Fe intersite charge transfer / Fe charge disproportionation with lattice coupling. | [18] |
| 38 | $SmCu_3Fe_4O_{12}$ | $SmFe_4(CuO_4)_3$ | 310 | A-site ordered ferrite; intersite charge transfer | Cu–Fe intersite charge transfer / Fe charge disproportionation with lattice coupling. | [18] |
| 39 | $YbCu_3Fe_4O_{12}$ | $YbFe_4(CuO_4)_3$ | 320.3 | A-site ordered ferrite; intersite charge transfer | Cu–Fe intersite charge transfer / Fe charge disproportionation with lattice coupling. | [18] |
| 40 | $Sr_3Fe_2O_7$ | $Sr_3Fe_2O_7$ | 340 | Layered ferrate; charge disproportionation | Charge-disproportionation transition in layered ferrate with Fe–O bond disproportionation and magnetic/lattice coupling. | [27] |
| 41 | $VO_2$ | $VO_2$ | 340 | Mott–Peierls/lattice-coupled transition | Rutile-to-monoclinic transition involving V–V dimerization, Peierls-like lattice instability, and strong electronic correlations. | [14] |
| 42 | $NiS_{1.48}Se_{0.52}$ / NiSSe | NiSeS | 350 | Bandwidth-controlled Mott transition | NiS–NiSe alloy transition associated with Mott physics in the antiferromagnetic phase and bandwidth/composition control. | [28] |
| 43 | $Sr_{1.2}La_{4.8}Mn_6O_{18}$ | $Sr_{1.2}La_{4.8}Mn_6O_{18}$ | 350 | A-site ordered manganite; dimensionality-driven | Dimensionality-driven insulator–metal transition in A-site excess perovskite manganite with Mn spin/charge/lattice coupling. | [22] |
| 44 | $Ca_2RuO_4$ | $Ca_2RuO_4$ | 357 | Layered ruthenate; Mott/structural transition | Mott insulating state controlled by $RuO_6$ octahedral distortion, orbital polarization, and structural reconstruction. | [29] |

*Continued on next page*

**Table 1: Supplementary Table S1 — Continued.**

| No. | MIT compound | Dataset formula | $T_{MIT}$ (K) | Representative mechanism class | Dominant or reported transition physics | Reference key(s) |
|---|---|---|---|---|---|---|
| 45 | $LuCu_3Fe_4O_{12}$ | $LuFe_4(CuO_4)_3$ | 367.2 | A-site ordered ferrite; intersite charge transfer | Cu–Fe intersite charge transfer / Fe charge disproportionation with lattice coupling. | [18] |
| 46 | $SmNiO_3$ | $SmNiO_3$ | 403 | Rare-earth nickelate; charge-transfer/bond disproportionation | Charge-transfer nickelate MIT with Ni–O bond disproportionation and magnetic/electronic reconstruction. | [12, 13, 30] |
| 47 | $NaOsO_3$ | $NaOsO_3$ | 410 | Slater-like AFM transition | Continuous antiferromagnetic MIT commonly described as Slater-like, with magnetic order opening/reorganizing the electronic gap. | [31] |
| 48 | h-FeS | FeS | 416.8 | Magnetically driven phonon instability | MIT enabled by magnetic order coupled to a phonon/lattice instability in hexagonal FeS. | [32, 33] |
| 49 | $V_3O_5$ | $V_3O_5$ | 425.2 | Magnéli vanadium oxide; mixed-valence localization | Mixed-valence vanadium oxide transition with electron localization, magnetic effects, and structural/electronic reconstruction. | [5, 11] |
| 50 | $BiCu_3Fe_4O_{12}$ | $Fe_4Cu_3BiO_{12}$ | 432 | A-site ordered ferrite; intersite charge transfer | Intermetallic/intersite charge transfer in Bi/La Cu–Fe quadruple perovskites with Fe valence rearrangement. | [18, 25] |
| 51 | $Ti_3O_5$ | $Ti_3O_5$ | 450 | Titanium oxide; polymorphic structural/electronic transition | MIT tied to temperature-driven polymorphic transition, Ti–Ti bonding rearrangement, and correlation/lattice coupling. | [34] |
| 52 | $EuNiO_3$ | $EuNiO_3$ | 460 | Rare-earth nickelate; charge-transfer/bond disproportionation | Charge-transfer nickelate MIT with Ni–O bond disproportionation and magnetic/electronic reconstruction. | [12, 30] |
| 53 | $Ti_2O_3$ | $Ti_2O_3$ | 466.5 | Corundum titanate; correlation + Ti–Ti bonding | Correlated transition/crossover involving Ti–Ti bonding, structural change, and electron localization. | [14, 35] |
| 54 | $LaCoO_3$ | $LaCoO_3$ | 500 | Cobaltite; spin-state/electronic crossover | Thermally driven spin-state crossover in $Co^{3+}$ coupled to bandwidth and charge-transfer gap; often not a sharp canonical MIT. | [36] |
| 55 | $GdNiO_3$ | $GdNiO_3$ | 509 | Rare-earth nickelate; charge-transfer/bond disproportionation | Small-R nickelate MIT with charge disproportionation, breathing distortion, and antiferromagnetic/electronic reconstruction. | [30, 37] |

*Continued on next page*

Table 1: Supplementary Table S1 — Continued.

| No. | MIT compound | Dataset formula | $T_{MIT}$ (K) | Representative mechanism class | Dominant or reported transition physics | Reference key(s) |
|---|---|---|---|---|---|---|
| 56 | $DyNiO_3$ | $DyNiO_3$ | 562 | Rare-earth nickelate; charge-transfer/bond disproportionation | Small-R nickelate MIT with charge disproportionation, breathing distortion, and antiferromagnetic/electronic reconstruction. | [30, 37] |
| 57 | $YNiO_3$ | $YNiO_3$ | 579 | Rare-earth nickelate; charge-transfer/bond disproportionation | Small-R nickelate MIT with charge disproportionation, breathing distortion, and antiferromagnetic/electronic reconstruction. | [30, 37] |
| 58 | $ErNiO_3$ | $ErNiO_3$ | 583 | Rare-earth nickelate; charge-transfer/bond disproportionation | Small-R nickelate MIT with charge disproportionation, breathing distortion, and antiferromagnetic/electronic reconstruction. | [30, 37] |
| 59 | $LuNiO_3$ | $LuNiO_3$ | 598.6 | Rare-earth nickelate; charge-transfer/bond disproportionation | Small-R nickelate MIT with charge disproportionation, breathing distortion, and antiferromagnetic/electronic reconstruction. | [30, 37] |
| 60 | $PrCoO_3$ | $PrCoO_3$ | 605 | Cobaltite; spin-state/electronic crossover | Thermally activated spin-state/electronic crossover involving $Co^{3+}$ states, bandwidth, and charge-transfer physics. | [36] |
| 61 | $NdCoO_3$ | $NdCoO_3$ | 635 | Cobaltite; spin-state/electronic crossover | Thermally activated spin-state/electronic crossover involving $Co^{3+}$ states, bandwidth, and charge-transfer physics. | [36] |
| 62 | $SmCoO_3$ | $SmCoO_3$ | 693 | Cobaltite; spin-state/electronic crossover | Thermally activated spin-state/electronic crossover involving $Co^{3+}$ states, bandwidth, and charge-transfer physics. | [36] |
| 63 | $GdCoO_3$ | $GdCoO_3$ | 740 | Cobaltite; spin-state/electronic crossover | Thermally activated spin-state/electronic crossover involving $Co^{3+}$ states, bandwidth, and charge-transfer physics. | [36] |
| 64 | $DyCoO_3$ | $DyCoO_3$ | 785 | Cobaltite; spin-state/electronic crossover | Thermally activated spin-state/electronic crossover involving $Co^{3+}$ states, bandwidth, and charge-transfer physics. | [36] |
| 65 | $YCoO_3$ | $YCoO_3$ | 800 | Cobaltite; spin-state/electronic crossover | Thermally activated spin-state/electronic crossover involving $Co^{3+}$ states, bandwidth, and charge-transfer physics. | [36] |

*Continued on next page*

**Table 1: Supplementary Table S1 — Continued.**

| No. | MIT compound | Dataset formula | $T_{\mathrm{MIT}}$ (K) | Representative mechanism class | Dominant or reported transition physics | Reference key(s) |
|---|---|---|---|---|---|---|
| 66 | $NbO_2$ | $NbO_2$ | 1070 | Peierls/Mott rutile-type transition | Rutile-related transition with Nb–Nb pairing/dimerization, lattice distortion, and correlation-assisted gap opening. | [14] |
| 67 | $BiFeO_3$ | $FeBiO_3$ | 1203 | High-temperature structural/electronic transition | Dataset entry corresponds to $BiFeO_3$; high-temperature beta/gamma structural transition with electronic-conductivity change, not a conventional low-temperature correlated MIT. | [38] |
| 68 | $TbNiO_3$ | $TbNiO_3$ | Not reported in processed summary | Rare-earth nickelate; charge-transfer/bond disproportionation | Small-R nickelate expected to follow charge-transfer/bond-disproportionation physics; no $T_{MIT}$ value was present in the processed summary used here. | [30, 37] |